\documentclass[sigconf]{acmart}

\usepackage{subcaption}
\usepackage{hyperref}
\usepackage{algorithm}
\usepackage{algpseudocode}
\usepackage{amsmath}
\usepackage{fontspec}
\usepackage{xspace}
\usepackage{tikz}
\usetikzlibrary{arrows.meta, positioning, calc, fit}
\usepackage{tcolorbox}
\usepackage{soul}
\usepackage{enumitem}

\newcommand{\head}[1]{\noindent\textbf{#1.}}

\newcommand{\tool}{\textsc{GATAS}\xspace} %
\newcommand{\wv}{\textsc{Waveform}\xspace} %
\newcommand{\pgd}{\textsc{PGD}\xspace}
\newcommand{\smack}{\textsc{SMACK}\xspace}

\renewcommand*{\equationautorefname}{Equation}
\def\equationautorefname~#1\null{(#1)\null}

\begin{document}

\title{Generative Testing of Automated Speech Recognition Systems}

\author{Yanis Xabier Wilbrand Peña}
\orcid{0009-0009-0980-8886}
\email{
}
\affiliation{%
  \institution{Technical University of Munich}
   \country{Germany}
}

\author{Oliver Wei{\ss}l}
\orcid{0009-0008-7575-0187}
\email{o.weissl@tum.de}
\affiliation{%
     \institution{Technical University of Munich}
   \country{Germany}
}

\author{Andrea Stocco}
\orcid{0000-0001-8956-3894}
\email{andrea.stocco@tum.de}
\affiliation{%
  \institution{Technical University of Munich \& fortiss GmbH}
  \country{Germany}
}

\renewcommand{\shortauthors}{Trovato et al.}

\settopmatter{printacmref=false} 
\renewcommand\footnotetextcopyrightpermission[1]{} 
\setcopyright{none}

\begin{abstract}
Automatic speech recognition (ASR) systems have achieved high accuracy with transformer-based models, enabling deployment in critical applications. However, they remain vulnerable to adversarial manipulation, particularly in black-box settings where attacks must preserve perceptual naturalness. This work introduces \tool, a black-box testing approach that generates failure inducing inputs by operating in the phoneme-level latent space of a text-to-speech model. Instead of perturbing waveforms directly, the approach interpolates latent representations to induce transcription errors while remaining within the manifold of natural speech. The attack is formulated as a multi-objective optimization problem balancing semantic divergence and perceptual quality. Our empirical evaluation against both white-box and black-box baselines shows that \tool achieves a 98\% success rate while producing lower distortion and higher perceptual quality, as confirmed by human studies. Despite operating without gradient access, \tool remains competitive against white-box methods, highlighting that representation and perceptual alignment are more critical than access to model internals. Overall, our results demonstrate that untargeted latent-space optimization enables the efficient generation of realistic and effective test cases for ASR systems.
\end{abstract} 






\maketitle

\section{Introduction}

Automatic speech recognition (ASR) has become a central interface between humans and machines, enabling applications such as voice assistants, transcription services, and accessibility tools~\cite{AHLAWAT2025201}. Recent transformer-based models, including Whisper~\cite{radford2023robust}, have reached performance levels that support deployment in high-stakes settings such as medical documentation, legal proceedings, and security-critical environments~\cite{KOKINA2025100734,Sloane2026}. This increased reliance demands systematic testing, as learning-enabled systems such as ASR are vulnerable to subtle input variations that can lead to incorrect behavior~\cite{2020-Humbatova-ICSE,2020-Riccio-EMSE}. In particular, adversarial manipulation has shown that benign-sounding speech can be modified to induce incorrect transcriptions without introducing perceptible artifacts, raising the question of whether ASR systems behave reliably not only under nominal conditions, but also under realistic, hard-to-detect variations of valid inputs.

Prior work has demonstrated the feasibility of adversarial attacks on ASR systems under both white-box and black-box assumptions, but with important limitations. White-box approaches~\cite{carlini2018audio} rely on gradient access to the target model, which restricts their applicability in real-world deployments. Black-box methods typically operate in waveform space, where perturbations tend to introduce audible distortions~\cite{alzantot2018did,taori2019targeted}. Transfer-based approaches avoid iterative querying but require training surrogate models~\cite{chen2020devil}, and the resulting adversarial audio still contains perceptible artifacts. In both cases, audio quality is treated as a secondary concern rather than an explicit optimization objective. While black-box attacks on ASR are well studied, methods that generate audio preserving both similarity to the original utterance and naturalness while inducing semantic divergence in the transcription remain limited. This gap motivates testing approaches that operate within a structured representation space rather than applying unconstrained perturbations to the raw audio signal.

To address this gap, this work proposes \tool (\textbf{G}ener\textbf{a}tive \textbf{T}esting of \textbf{A}utomatic Speech Recognition \textbf{S}ystems), a novel black-box testing approach for ASR systems that generates natural-sounding audio while altering transcription semantics. \tool leverages a text-to-speech (TTS) model, specifically StyleTTS2~\cite{li2023styletts}, which represents speech as phoneme-level embedding vectors encoding acoustic properties. Instead of modifying the waveform directly, the approach interpolates between the ground-truth phoneme embeddings and a noise reference, with a per-phoneme scalar weight controlling the degree of perturbation. This shifts individual phoneme realizations while the TTS decoder ensures that the generated audio remains natural-sounding. \tool formulates test generation as a multi-objective optimization problem that balances semantic divergence against perturbation imperceptibility, enabling the discovery of test cases that induce meaningful transcription changes while preserving audio quality.

Our empirical study uses the Harvard Sentences corpus~\cite{rothauser1969ieee} and targets Whisper~\cite{radford2023robust}, a modern end-to-end encoder--decoder transformer-based ASR system. To assess the impact of the interpolation domain, we compare \tool against a set of representative baselines covering both black-box and white-box settings. In particular, we include a waveform-based variant that adopts the same multi-objective optimization setup as \tool but performs interpolation directly between the ground-truth audio and Gaussian noise, thereby isolating the choice of search space as the only experimental variable. We further consider \smack~\cite{yu2023smack}, a black-box method combining genetic search with gradient estimation over prosodic feature representations, which we adapt from its original targeted formulation to an untargeted setting. Finally, we include \pgd~\cite{olivier2022there}, a white-box gradient-based attack, as a reference point approximating the performance achievable with full access to model internals.
Our evaluation considers three aspects: (i)~effectiveness, measured by semantic divergence and perturbation imperceptibility; (ii)~validity, assessed through human evaluation of intelligibility and naturalness; and (iii)~efficiency, quantified by query budget and runtime.

The results of our empirical study show that embedding-space manipulation achieves high combined success in finding misbehavior inducing inputs while balancing semantic divergence and audio quality simultaneously. \tool achieves a success rate of 98\% compared to 99\% for the \wv baseline, 100\% for \smack, and 100\% for \pgd. The \wv baseline achieves semantic divergence only through signal destruction, and \smack, originally designed for pre-transformer ASR systems such as DeepSpeech2~\cite{amodei2016deep}, exhibits similarly degraded performance on Whisper. \pgd achieves comparable audio quality to the proposed method, as expected given full gradient access, but requires white-box access unavailable in practical deployment. Human evaluation confirms these findings: \tool achieves mean technical quality ratings of 4.81 vs.\ 4.92 for ground truth, and mean semantic quality ratings of 3.86 vs.\ 4.01 for ground truth, while the other baselines receives from 1.24-3.73 on technical quality and 1.33-3.98 on semantic quality. The cost per successful test case is approximately 40\% lower than the \wv baseline and for \smack the cost is nearly identical, while querying the ASR model nearly 17 times as often, showing the efficiency per query.

Our paper makes the following contributions:

\begin{description}[noitemsep,nolistsep]

\item [Approach] We propose \tool, a black-box testing approach for ASR systems that operates in the phoneme-level embedding space of a TTS model. By interpolating between ground-truth and noise embeddings at the per-phoneme level, the method generates natural-sounding audio while inducing controlled semantic changes in ASR transcriptions.

\item [Methodology] We formulate test generation as a multi-objective optimization problem balancing semantic divergence and perturbation imperceptibility. This enables efficient discovery of test cases that alter ASR transcriptions while preserving audio similarity and naturalness.

\item [Empirical Study] We conduct an empirical evaluation on a standard speech benchmark and a transformer-based ASR system, comparing against both black-box and white-box baselines. The results show that \tool achieves effectiveness comparable to the white-box baseline despite requiring only query access, while substantially outperforming the black-box baselines in audio quality and similarity to the original speech. Human evaluation confirms that the generated audio remains intelligible and natural-sounding.

\end{description}

\section{Background}

\begin{figure*}[t]
    \centering
    \begin{tikzpicture}[
        block/.style={rectangle, draw, rounded corners, minimum width=2.9cm, minimum height=1.0cm, align=center},
        arrow/.style={-Latex, thick},
        node distance=1.6cm and 1.4cm
    ]

\node[block] (tts) {TTS Model\\\textit{(StyleTTS2)}};
\node[block, below=0.55cm of tts] (interp) {Interpolation};

\node[draw, dashed, thick, rounded corners, fit=(interp)(tts),
      label={\tool-Manipulator}, inner sep=0.18cm] (manipbox) {};

\node[block, right=of tts] (sut) {SUT\\\textit{(Whisper)}};
\node[block, right=of sut] (objectives) {Objectives\\$F_\bullet$};

\node[block, right=of interp] (opt) {Optimizer\\\textit{(NSGA-II)}};

\draw[arrow] ($(manipbox.west)+(-2.3,0.55)$) -- ($(manipbox.west)+(0.0,0.55)$)
    node[pos=0.5, above] {initial text $t$};

\draw[arrow] ($(manipbox.west)+(-2.3,-0.55)$) -- ($(manipbox.west)+(0.0,-0.55)$)
    node[pos=0.5, below] {initial random $\kappa$};

\draw[arrow] (interp.north) -- (tts.south)
    node[pos=0.5, right] {$\mathbf{h}'_{\mathrm{text}}$};

\draw[arrow] (tts.east) -- (sut.west)
    node[pos=0.25, below] {$x$};

\draw[arrow] (sut.east) -- (objectives.west)
    node[pos=0.5, above] {$\hat{t}$};

\draw[arrow] (objectives.south) |- (opt.east)
    node[pos=0.25, right] {$\mathbb{R}$};

\draw[arrow] (opt.west) -- (interp.east)
    node[pos=0.5, below] {new $\kappa$};

\end{tikzpicture}
    \caption{Overview of \tool. The initial input text $t$ is processed by theManipulator, where interpolation between $\textbf{h}^{\mathrm{GT}}_{\text{text}}$ and $\textbf{h}_{\mathrm{noise}}$ parameterized by $\kappa$ produces $\textbf{h}'_{\mathrm{text}}$, which is synthesized by the TTS model into audio $x$. The SUT then produces transcription $\hat{t}$, which is evaluated by the objectives and used by the optimizer to update $\kappa$.}
    \label{fig:tool_loop}
\end{figure*}
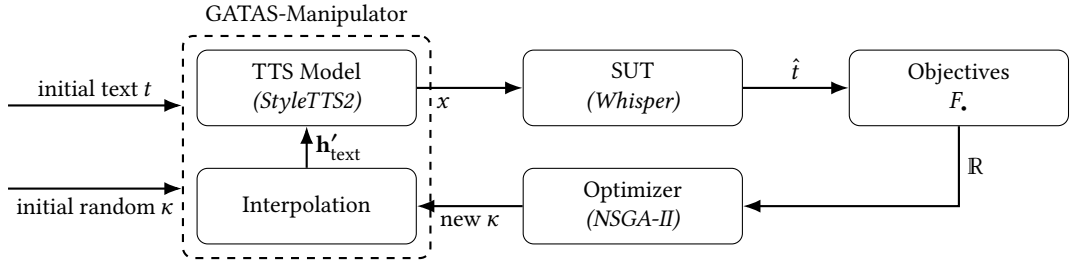

\subsection{Automatic Speech Recognition}

ASR systems map continuous audio signals $\in \mathcal{X}$ to discrete token sequences $\in \mathcal{T}$, $\operatorname{ASR}: \mathcal{X} \rightarrow \mathcal{T}$.

This requires modeling both acoustic-phonetic structure and linguistic constraints~\cite{rabiner1993fundamentals}. The input waveform is typically transformed into a log-mel spectrogram, which serves as a time–frequency representation for neural processing. Classical ASR systems combined Gaussian mixture models with Hidden Markov Models to align acoustic features to phoneme states, together with n-gram language models for decoding. Modern systems instead use end-to-end neural architectures~\cite{ahlawat2025automatic}. End-to-end models either learn implicit alignments~\cite{graves2006connectionist} or generate token sequences autoregressively~\cite{chan2015listen}. These approaches are commonly instantiated with transformer architectures~\cite{vaswani2017attention, radford2023robust}, whose self-attention mechanisms capture long-range dependencies and benefit from large-scale pre-training, significantly improving transcription accuracy.

\subsection{Text-to-speech}

TTS systems perform the inverse mapping, generating speech from textual input 
$\operatorname{TTS}: \mathcal{T} \rightarrow \mathcal{X}$.
Early pipelines separated text analysis, acoustic modeling, and waveform synthesis, often producing intelligible but unnatural speech. Neural TTS replaces these components with learned models. Sequence-to-sequence architectures such as Tacotron~\cite{wang2017tacotron} learn alignments between phonemes and acoustic frames, while neural vocoders such as HiFi-GAN~\cite{kong2020hifi} generate high-fidelity waveforms. Recent models operate in learned latent spaces, where speech is represented by compact vectors encoding content, speaker identity, and prosody~\cite{li2023styletts}. These representations enable direct manipulation of acoustic properties at a fine-grained level, which is central to the attack mechanism studied in this work.

\subsection{Testing Methods for ASR Systems}
The majority of literature on testing methods for ASR Systems comes from adversarial machine learning settings, where inputs are perturbed to induce incorrect model outputs. Prior work has shown that neural networks are sensitive to small, structured perturbations, giving rise to adversarial examples~\cite{szegedy2013intriguing,goodfellow2014explaining}. Such inputs are constructed through adversarial attacks, which are commonly categorized by the attacker's knowledge of the system (e.g., white-box, gray-box, black-box) and the attack objective (targeted vs.\ untargeted)~\cite{goodfellow2014explaining,carlini2017towards,weissl2025targeted}.

Adversarial testing primarily targets the robustness of deep learning, and thus automated speech recognition ASR systems, under worst-case perturbations. However, many real-world tasks require evaluating generalization, as training data cannot exhaustively capture the variability of real-world conditions. This creates a gap: existing testing methodologies emphasize robustness, while the systematic assessment of generalization remains underexplored.

Prior work in ASR testing reflects this limitation. Metamorphic testing approaches generate variants via affine transformations in waveform space, enabling robustness evaluation under controlled perturbations, but they do not produce functionally novel inputs; instead, they iteratively corrupt existing samples~\cite{ji2022asrtest}. In contrast, work such as AequeVox highlights that generalization is critical for fairness, showing that natural variations in speech—e.g., due to native language or voice characteristics—can lead to significant performance degradation~\cite{rajan2022aequevox}.

Testing generalization therefore requires the creation of functionally new inputs that extend beyond the training distribution while remaining realistic. This work addresses this gap by generating novel test cases using TTS models. The generation process is guided by constraints that enforce realism and effectiveness, enabling systematic exploration of previously unseen yet plausible inputs for evaluating ASR system behavior.




\subsection{Motivating Example}

To illustrate the challenges of testing ASR systems, consider this example. A user issues a simple voice command: \textit{``Turn on the light''}. A standard ASR system correctly transcribes the utterance under nominal conditions. However, small variations in speech, such as changes in pronunciation, prosody, or acoustic realization, can lead to incorrect transcriptions such as 
\textit{``Turn on the night''} or \textit{``Turn on the line''}. 
These variations may arise naturally (e.g., due to accent or speaking style) or be induced artificially. Importantly, such changes do not require large or unrealistic perturbations, but can result from subtle modifications in the acoustic signal that remain perceptually plausible.
Existing testing approaches primarily generate variants by perturbing existing recordings, limiting coverage to transformations of known inputs. In contrast, our approach leverages TTS models to generate new speech samples that systematically explore variations in phonetic realization while preserving naturalness. This enables the construction of new functional test cases that remain realistic yet expose previously unseen failures.

\section{Methodology}\label{sec:methodology}

This work proposes \tool, a novel black-box test case generator for ASR systems. The pipeline is illustrated in \autoref{fig:tool_loop} and comprises four components: (i)~the \textit{System under Test (SUT)}, namely the target ASR system, (ii)~a \textit{Manipulator}, which perturbs the phoneme-level acoustic latent space of a TTS model to generate candidate audio samples, (iii)~an \textit{Optimizer} to find optimal solutions, and (iv)~\textit{Objectives} that evaluate semantic transcription divergence and perturbation imperceptibility. 

A key design choice of our approach is to operate in the latent space of a TTS model rather than directly in the waveform domain. This choice fundamentally affects the structure of the optimization problem. Indeed, in waveform space, inputs are high-dimensional and lack semantic structure: small perturbations do not correspond to meaningful changes in phonetic content, and optimization tends to introduce noise-like artifacts. As a result, semantic divergence and perceptual quality become conflicting objectives, where increasing one degrades the other.

In contrast, the TTS latent space is organized around acoustic meaning. Interpolations correspond to controlled variations in phoneme realization, and the generative decoder constrains outputs to remain on the manifold of natural speech. This induces a well-structured search space in which semantic divergence and perceptual quality become compatible objectives, enabling effective and realistic test generation.

\begin{algorithm}[t]
\caption{\tool}
\label{alg:attack}
\begin{algorithmic}[1]
\Require Ground truth text $\text{GT}$, black-box ASR system
\Ensure Generated audio
\State Extract $\mathbf{h}_\text{text}^\text{GT}$ from \textit{TTS} given $\text{GT}$
\State Sample $\mathbf{h}_\text{noise} \sim \mathcal{N}(\mu_\text{GT}, \sigma_\text{GT})$
\State Initialize population $P$, archive $A \leftarrow \emptyset$
\For{generation $g = 1$ to $\text{max\_generations}$}
    \For{each individual $\kappa$ in $P$}
        \State $\mathbf{h}_\text{text}' \leftarrow (1 - \kappa) \odot \mathbf{h}_\text{text}^\text{GT} + \kappa \odot \mathbf{h}_\text{noise}$
        \State $\text{audio} \leftarrow \text{\textit{TTS}}(\mathbf{h}_\text{text}',\ \text{other features from GT})$
        \State $\text{transcription} \leftarrow \text{\textit{SUT}}(\text{audio})$
        \State $F_1 \leftarrow \text{SetOverlap}(\text{transcription},\ \text{GT})$
        \State $F_2 \leftarrow \text{PESQ}(\text{audio},\ \text{audio}_\text{GT})$
        \State $\text{fitness}[\kappa] \leftarrow (F_1, F_2)$
    \EndFor
    \State $A \leftarrow \text{non\_dominated}(A \cup P)$
    \If{any $\kappa$ in $A$ early stopping criteria}
        \State \textbf{break}
    \EndIf
    \State $P \leftarrow \text{\textit{Optimizer}}(P \cup A,\ \text{fitness})$
\EndFor
\State $\kappa_\text{best} \leftarrow \text{threshold\_proportional\_l3\_selection}(A)$
\State $\mathbf{h}_\text{text}' \leftarrow (1 - \kappa_\text{best}) \odot \mathbf{h}_\text{text}^\text{GT} + \kappa_\text{best} \odot \mathbf{h}_\text{noise}$
\State \Return $\mathit{TTS}(\mathbf{h}_\text{text}, \text{other features from GT})$
\end{algorithmic}
\end{algorithm}

\autoref{alg:attack} details the test generation process. First, the ground-truth acoustic embedding $\mathbf{h}_{\text{text}}^{\text{GT}}$ is extracted from the TTS encoder (line~1), a noise embedding $\mathbf{h}_{\text{noise}}$ is sampled as a directional reference (line~2), and a population of candidate solutions is initialized  (line~3). Second, in the optimization loop, each candidate, representing interpolation weights for the embeddings, results in a perturbed embedding $\mathbf{h}_{\text{text}}'$ which is synthesized by the TTS model into audio and evaluated by the ASR system (lines~5--12). The resulting transcription and perceptual quality are used to guide the search  (line~17). Third, from all non-dominated solutions encountered during optimization, the best candidate is selected for the final solution (lines~20-21).

In the remaining of the section, we provide additional details on each component of \tool.

\subsection{System under Test}\label{sec:sut}

The system under test is an ASR system that maps an input audio signal \( x \) to a transcription \( t = f(x) \). In this work we consider a strict black-box setting in which no access to system internals is available, including architecture, parameters, gradients, or intermediate outputs. In short, the only observable output is the final transcription produced by the system.

We focus on untargeted test generation, where any semantically meaningful deviation from the original transcription constitutes success, because this formulation reflects realistic scenarios, where specifying a target transcription is impractical due to the open vocabulary and variable-length outputs of ASR systems.

Our approach operates in a digital setting, where generated audio is directly provided to the ASR system. A test case is considered failure inducing when it induces a semantic change in the transcription while preserving perceptual plausibility of the audio.


\subsection{Manipulator}\label{sec:manipulator}

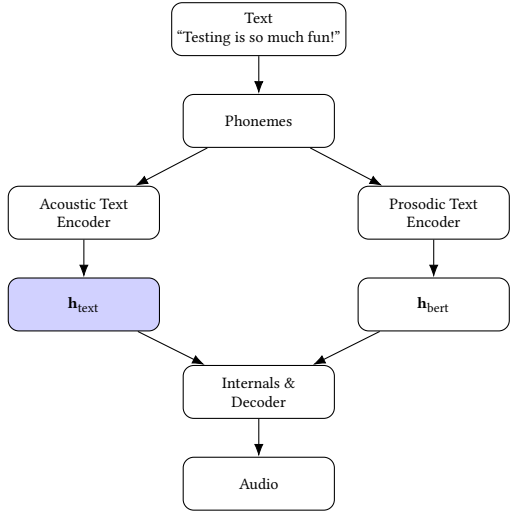
\begin{figure}
\centering
\begin{tikzpicture}[
    >=Latex,
    node distance=0.5cm and 0.3cm,
    every node/.style={font=\scriptsize},
    box/.style={
        draw,
        rounded corners,
        minimum width=2.0cm,
        minimum height=0.7cm,
        align=center,
        inner sep=2pt
    },
    imp/.style={
        box,
        fill=blue!18
    }
]

\node[box] (text) {Text\\{\scriptsize``Testing is so much fun!''}};
\node[box, below=of text] (token) {Phonemes};

\node[box, below left=of token] (acoustic) {Acoustic Text\\Encoder};
\node[box, below right=of token] (prosodic) {Prosodic Text\\Encoder};

\node[imp, below=of acoustic] (htext) {$\mathbf{h}_{\text{text}}$};
\node[box, below=of prosodic] (hbert) {$\mathbf{h}_{\text{bert}}$};

\node[box, below=0.8cm of $(htext)!0.5!(hbert)$] (internals) {Internals \&\\Decoder};

\node[box, below=of internals] (audio) {Audio};

\draw[->] (text) -- (token);
\draw[->] (token) -- (acoustic);
\draw[->] (token) -- (prosodic);

\draw[->] (acoustic) -- (htext);
\draw[->] (prosodic) -- (hbert);

\draw[->] (htext) -- (internals);
\draw[->] (hbert) -- (internals);

\draw[->] (internals) -- (audio);

\end{tikzpicture}
\caption{Simplified StyleTTS2 pipeline~\cite{li2023styletts}.}
\label{fig:tts}
\end{figure}

The manipulation operates in the phoneme-level acoustic embedding space of a TTS model. Specifically, \tool uses the acoustic text embeddings $\mathbf{h}{\text{text}} \in \mathbb{R}^{p \times 512}$ produced by the StyleTTS2 acoustic encoder~\cite{li2023styletts}, as shown in \autoref{fig:tts}. This representation encodes the phonetic content of the synthesized speech, while other components (e.g., prosody and style) govern timing, expressiveness, and speaker characteristics. By modifying $\mathbf{h}_{\text{text}}$, \tool alters the phonetic content while preserving these attributes.
Importantly, operating in this latent space constrains generated audio to remain on the natural speech manifold, as enforced by the TTS decoder.

We choose to manipulate $\mathbf{h}_{\text{text}}$ as alternative intervention points are less suitable. Style variables primarily encode speaker identity and exhibit limited variability in our setting (\autoref{fig:tts} Internals). Prosodic representations mainly control timing and rhythm, and perturbing them degrades naturalness without reliably affecting phonetic content (found in \autoref{fig:tts} under $\mathbf{h}_{\operatorname{bert}}$). Downstream representations are already temporally expanded or entangled with style information, making controlled manipulation more difficult. In contrast, $\mathbf{h}_{\text{text}}$ provides a compact and semantically aligned representation for phoneme-level control.

Let $\mathbf{h}_{\text{text}}^{\text{GT}}$ denote the embedding extracted from the ground-truth input. Direct optimization over $\mathbf{h}_{\text{text}} \in \mathbb{R}^{p \times 512}$ is intractable due to its dimensionality ($512p$ degrees of freedom). We therefore parameterize the manipulation using a per-phoneme interpolation vector

\begin{equation}
\kappa \in [0,1]^p,
\end{equation}
where each $\kappa_i$ controls the degree of perturbation applied to the $i$-th phoneme.
To define a reference direction in latent space, we sample a noise embedding matched to the statistics of the ground-truth representation:
\begin{equation}
\mathbf{h}_{\text{noise}} \sim \mathcal{N}(\mu_{\text{GT}}, \sigma_{\text{GT}}),
\end{equation}
where $\mu_{\text{GT}}$ and $\sigma_{\text{GT}}$ are computed element-wise from $\mathbf{h}_{\text{text}}^{\text{GT}}$.
The perturbed embedding is then computed as
\begin{equation}
\mathbf{h}_{\text{text}}' = (1 - \kappa) \odot \mathbf{h}_{\text{text}}^{\text{GT}} + \kappa \odot \mathbf{h}_{\text{noise}},
\end{equation}
where $\kappa$ is broadcast using the Hadamard product $\odot$ across the 512-dimensional phoneme embeddings. This reduces the search space from $512p$ to $p$ dimensions while enabling fine-grained control over individual phonemes.

Finally, the modified embedding $\mathbf{h}_{\text{text}}'$ is passed through the TTS model to synthesize the new audio for the proposed test case.





\subsection{Optimizer}\label{sec:optimizer}

We formulate test generation as a multi-objective optimization problem~\cite{deb2002fast}, using a gradient-free evolutionary algorithm that maintains a diverse set of Pareto-optimal solutions. The objective is to optimize a set of objectives capturing semantic and perceptual properties of the generated audio.

The population is initialized by sampling individuals uniformly from $[0,1]^p$, augmented with a seeded individual $\sum\kappa=\mathbf{1}$ to ensure coverage of high-divergence regions of the search space.

At each generation, individuals are evaluated with respect to the objective set and ranked via non-dominated sorting, with selection performed using crowding distance to preserve diversity.
To retain high-quality solutions across generations, we maintain an archive of non-dominated individuals:
\begin{equation}
A \leftarrow \text{non\_dominated}(A \cup P).
\end{equation}

The optimization proceeds until either the query budget is exhausted or an early stopping criterion is met. The early stopping condition is defined in terms of the objectives and is detailed in the next section. The final solution is selected from the archive based on a balanced trade-off across the objective set.

\subsection{Objectives}\label{sec:objectives}

The objectives of \tool test generation aim to balance semantic divergence of the ASR output and perceptual similarity of the generated audio. In particular, two objectives are used.  

\head{Semantic Divergence (SetOverlap)}\label{sec:setoverlap}
The first objective measures semantic preservation using a set-based overlap between content words in the ground truth transcription and the ASR output. To ensure that only semantically meaningful differences are considered, both texts undergo identical pre-processing.

First, function words (e.g., ``the'', ``is'', ``on'') are removed using a standard English stop-word list $\mathcal{S}$~\cite{bird2009natural}, as these contribute little semantic content and are typically robust to perturbations. Second, remaining words are normalized via WordNet lemmatization~\cite{miller1995wordnet}, mapping each word $w$ to its canonical base form $\lambda(w)$. This collapses morphological variants (e.g., ``sliding'' $\rightarrow$ ``slide'', ``bananas'' $\rightarrow$ ``banana'') into a single representation.
The resulting content-word set is defined as:
\begin{equation}
W(\text{text}) = \{\, \lambda(w) \mid w \in \text{words}(\text{text}),\ w \notin \mathcal{S} \,\}
\end{equation}

Then, SetOverlap is computed as:
\begin{equation}
F_1 = \frac{|W(\text{GT}) \cap W(\text{ASR})|}{|W(\text{GT})|}
\end{equation}

This metric measures the fraction of original semantic content preserved in the ASR transcription. A value of $0.0$ indicates complete semantic divergence, while $1.0$ indicates perfect preservation.

\head{Perceptual Similarity (PESQ)}
Perceptual similarity between the test case and ground-truth audio is measured using PESQ~\cite{rix2001perceptual}. PESQ produces a raw score in the range $[-0.5, 4.5]$, which correlates with human perceptual judgments and can be mapped to the MOS-LQO scale~\cite{itu1996mos}.
To use PESQ as a minimization objective, we normalize and invert the score:
\begin{equation}
\text{PESQ}_{\text{norm}} = \frac{\text{PESQ}_{\text{raw}} + 0.5}{5.0}, \quad
F_2 = 1 - \text{PESQ}_{\text{norm}}
\end{equation}

Lower values of $F_2$ correspond to higher perceptual quality, i.e., closer similarity to the original audio.

\section{Empirical Study}
\label{sec:empirical}

\subsection{Research Questions}
\label{sec:research-questions}

We investigate three research questions:

\textbf{RQ\textsubscript{1} (effectiveness):} \textit{How effective is \tool in generating audio test cases that induces semantic divergence in ASR transcriptions while preserving perturbation imperceptibility? }

\textbf{RQ\textsubscript{2} (validity):} \textit{Are the generated test cases by \tool perceived by humans as natural and intelligible?}

\textbf{RQ\textsubscript{3} (efficiency):} \textit{How efficient is \tool in generating failure inducing test inputs?}

RQ\textsubscript{1} evaluates the ability of our approach to achieve meaningful transcription changes and high audio similarity. In addition to overall success, we analyze semantic divergence and audio quality separately to assess whether the generated test cases induce controlled and non-trivial changes rather than noisy outputs.

RQ\textsubscript{2} assesses whether the generated test cases remain perceptually valid by evaluating human judgments of audio quality and intelligibility. This determines whether the generated test cases are realistic speech signals rather than degenerate artifacts.

RQ\textsubscript{3} measures efficiency in terms of ASR query usage, convergence speed, and runtime, and compare these aspects against our baseline methods to determine whether the approach remains computationally practical.

\subsection{Metrics}\label{sec:metrics}

\subsubsection{RQ\textsubscript{1}}

To evaluate generated test cases, we consider multiple complementary dimensions capturing behavioral impact, perceptual realism, and perturbation magnitude. These dimensions are necessary, as changes in transcription alone do not indicate meaningful failures, and perceptual plausibility is required to ensure realistic test inputs, using the following metrics.

\head{Success} ($\uparrow$): Measures the fraction of generated test cases that change the ASR transcription, indicating altered system behavior. As transcription changes are not inherently erroneous, this metric is evaluated alongside measures that quantify the magnitude and nature of these changes.

\head{UTMOS} ($\uparrow$): A neural approximation of the mean opinion score for audio naturalness~\cite{saeki2022utmos}. This metric assesses whether generated test cases remain perceptually plausible, as unrealistic or nonsensical audio would invalidate expected ASR behavior.

\head{Spectrogram-$d_F$} ($\downarrow$): Quantifies perturbation magnitude using the Frobenius distance between MEL-spectrograms, capturing changes in time–frequency representations. Lower values indicate less destructive and more subtle modifications.

\head{Raw-$d_C$} ($\downarrow$): Measures character-level differences in ASR outputs, capturing fine-grained changes such as insertions, deletions, or substitutions that may not significantly affect semantic meaning.

\head{Embedding-$d_C$} ($\downarrow$): Assesses semantic change by computing distances between sentence embeddings of ASR outputs using the \textit{all-MiniLM-L6-v2} model from the \textit{sentence-transformers} library. This captures higher-level content shifts while remaining reproducible and computationally efficient.

\subsubsection{RQ\textsubscript{2}}

To assess the perceptual quality of generated test cases, we conduct a human evaluation focusing on both audio fidelity and semantic intelligibility. We report Mean Opinion Score (MOS) ratings along two complementary dimensions.

\head{Technical MOS} ($\uparrow$): Assesses perceptual audio quality in terms of clarity, distortion, and noise. Ratings are collected using the MOS scale~\cite{itu_p10}, ranging from 1 (heavily corrupted) to 5 (perfectly clear). Participants respond to the question: \textit{How would you rate the technical audio quality (e.g., noise, distortion, clarity)?}

\head{Semantic MOS} ($\uparrow$): Evaluates the perceived intelligibility and coherence of the spoken content. Using the same MOS scale (1--5), ratings range from 1 (nonsensical or unintelligible) to 5 (fully coherent and meaningful). Participants respond to the question: \textit{How would you rate the semantic quality of the spoken content?}

\subsubsection{RQ\textsubscript{3}}

To evaluate the efficiency of test generation, we measure both computational cost and interaction requirements with the system under test, using the following metrics.

\head{Runtime per attack} ($\downarrow$): Measures the time required to generate a single test case, reported in seconds. Although runtime depends on implementation details and hardware, this metric provides a comparable estimate of computational cost across methods.

\head{Query budget} ($\downarrow$): Captures the number of interactions with the system under test (SUT) during test case generation. This metric must be interpreted with caution: white-box methods require fewer queries due to access to internal parameters, and different black-box methods rely on fundamentally different optimization strategies, preventing direct alignment across methods.

\subsection{Objects of Study}\label{sec:objects}

\subsubsection{Dataset}

We use the Harvard Sentences corpus~\cite{rothauser1969ieee}, which consists of 720 phonetically balanced English sentences originally designed for speech intelligibility evaluation.\footnote{\url{https://www.cs.columbia.edu/~hgs/audio/harvard.html}} We select the first 100 sentences to evaluate the proposed TTS-based attack and the baselines. The dataset is well suited to our setting: its phonetic balance ensures coverage of diverse acoustic conditions, while low semantic predictability reduces reliance on language-model priors, forcing ASR decisions to be driven primarily by acoustic input.

\subsubsection{System under Test}

We adopt Whisper tiny (\texttt{openai/whisper-tiny})~\cite{radford2023robust}, an end-to-end encoder-decoder transformer trained on 680{,}000 hours of diverse, multilingual, and noisy audio data. The ASR system is treated strictly as a black box: only final transcriptions are observed, with no access to gradients, probabilities, or activations.

Decoding temperature is set to 0, enforcing greedy decoding and deterministic outputs. This ensures identical audio inputs produce identical transcriptions, which is necessary for stable fitness evaluation. Whisper tiny is also chosen for computational efficiency, enabling more candidate evaluations within a fixed time budget—important for population-based optimization, where query count affects convergence.

\subsubsection{Baselines}

The \wv baseline uses the same evolutionary computation as \tool but operates directly in waveform space. Instead of manipulating latent representations, interpolation is applied between the TTS-generated ground-truth waveform and Gaussian noise. All other components, including objective functions, success criteria, and early stopping, remain identical. This isolates the effect of representation space, so performance differences are attributable to the search space rather than algorithmic factors.

We further compare against \smack~\cite{yu2023smack}, a black-box attack combining genetic search with gradient estimation over prosodic features. As \smack was originally designed for targeted attacks, we adapt it to the untargeted setting by replacing its loss function accordingly. Original optimization parameters are retained, as the methods use fundamentally different frameworks.

Finally, we include \pgd~\cite{olivier2022there}, a white-box baseline to assess performance relative to methods with access to model internals. As its optimization is not directly compatible with our setting, we use the parameters specified in the original work.

\subsection{Experimental Setup}\label{sec:setup}

For RQ\textsubscript{1}, we executed \tool and the baselines with a population size selected via preliminary experiments over $\{50, 100, 200\}$. Smaller population sizes led to premature convergence, while 100 provided stable and quick exploration and was therefore adopted. The number of generations is fixed to 100, yielding a maximum query budget of 10{,}000 evaluations per run. Early stopping is applied once both success thresholds are satisfied.

The mutation probability is defined per variable as $1/n_{\text{var}}$, adapting automatically to the dimensionality of the interpolation vector.
For early stopping, a run is considered successful when both objectives satisfy:
\begin{equation}
\text{SetOverlap} \leq 0.5 \quad \land \quad \text{PESQ}_{\text{norm}} \leq 0.2
\end{equation}

The SetOverlap threshold enforces that at least half of the semantically meaningful content words are changed, preventing trivial variations. The normalized PESQ threshold corresponds to a raw PESQ score of approximately $\geq 3.5$ (MOS-LQO $\geq 3.55$), indicating perceptual quality between fair and good~\cite{itu_p862_1_2003}. These thresholds balance semantic divergence and perceptual similarity and are aligned with prior work in testing audio consuming deep learning systems~\cite{itu_p862_1_2003}.

To answer RQ\textsubscript{2}, we conduct a human study to assess perceptual audio quality along two dimensions: technical quality and semantic quality. The evaluation covers five conditions: unperturbed ground truth audio synthesized by StyleTTS2, \tool, \wv, \smack, and \pgd. For each condition, 10 sentences are included, yielding 50 audio clips in total.

A total of 10 participants, recruited from an academic environment and indipedent from this work, each evaluated all 50 clips. For every audio clip, participants rated both technical and semantic quality on a 1 to 5 MOS scale, as defined in \autoref{sec:metrics}. In total, this yields $50 \times 10 = 500$ individual ratings per dimension.
\subsection{Results}

\subsubsection{Effectiveness (RQ\textsubscript{1})}\label{sec:rq1}

\begin{table}[t]
\caption{Effectiveness results (RQ\textsubscript{1}) for all approaches.}
\resizebox{\columnwidth}{!}{%
\begin{tabular}{llllll}\toprule
         & \textit{Success} $\uparrow$& UTMOS  $\uparrow$& $d_F$-Spec  $\downarrow$ & $d_C$-Raw  $\downarrow$& $d_C$-Emb $\downarrow$ \\ \midrule
\tool    & 0.980                   & \textbf{4.466 $\pm$ 0.030 }& \textbf{0.017 $\pm$ 0.002} & \textbf{0.047 $\pm$ 0.066} & \textbf{0.439 $\pm$ 0.190} \\
\wv & 0.990                  & 2.108 $\pm$ 0.108 & 0.106 $\pm$ 0.023 & 0.695 $\pm$ 0.269 & 0.908 $\pm$ 0.114 \\

\smack & \textbf{1.000}       & 2.101 $\pm$ 0.067& 0.270 $\pm$ 0.024 & 0.191 $\pm$ 0.177 & 0.806 $\pm$ 0.224 \\

\pgd      & \textbf{1.000}                   & 4.443 $\pm$ 0.041 & 0.045 $\pm$ 0.006 & 0.157 $\pm$ 0.103 & 0.870 $\pm$ 0.130\\

\bottomrule
\end{tabular}\label{tab:performance}}
\end{table}
\begin{table}[t]
\caption{Statistical comparison of baselines versus \tool. Wilcoxon signed-rank test p-values and Cohen's d effect sizes; \textbf{bold} indicates statistical significance.}
\resizebox{\columnwidth}{!}{%
\begin{tabular}{lcccccccc}\toprule
 & \multicolumn{2}{l}{UTMOS} & \multicolumn{2}{l}{$d_F$-Spec} & \multicolumn{2}{l}{$d_C$-Raw} & \multicolumn{2}{l}{$d_C$-Embmb} \\ \midrule
\wv & \textbf{1.94e-18}      & 20.4      & \textbf{1.95e-18}        & 3.8       &\textbf{ 2.13e-18 }       & 2.4       & \textbf{1.88e-17 }       & 2.1       \\
\smack & \textbf{1.95e-18}&  31 & \textbf{1.95e-18}& 10.4&\textbf{1.79e-16} &0.8 &\textbf{3.06e-15} & 1.3\\
\pgd      & \textbf{4.44e-03}      & 0.3       & \textbf{1.95e-18 }       & 5.1       &\textbf{ 1.84e-16}        & 0.9       & \textbf{2.05e-17}        & 1.7   \\  \bottomrule
\end{tabular}\label{tab:stats}}
\end{table}

\begin{figure*}[t]
\centering
\begin{subfigure}[t]{0.24\textwidth}
    \centering
    \includegraphics[width=\linewidth]{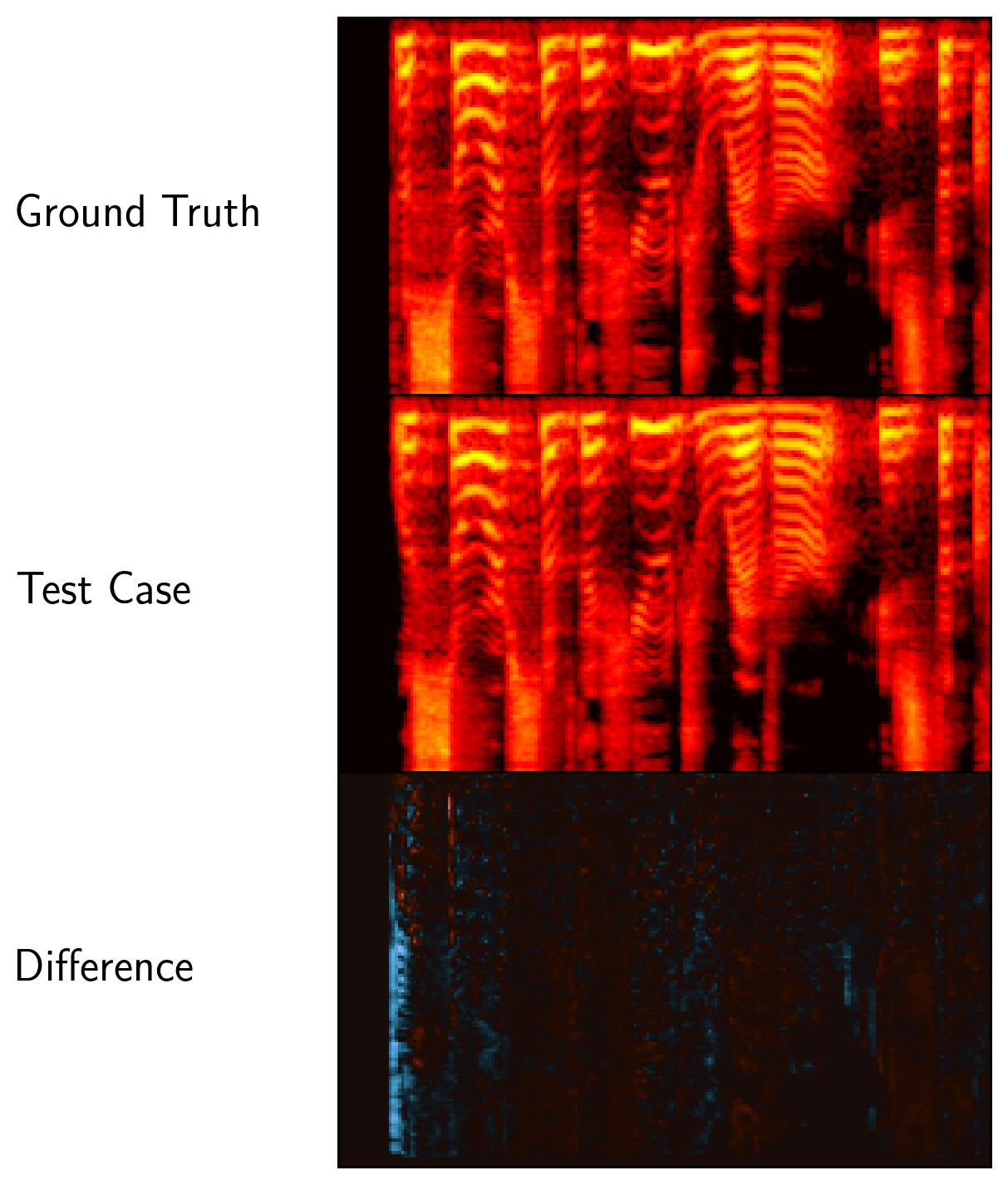}
    \caption{\tool}
\end{subfigure}
\hfill
\begin{subfigure}[t]{0.24\textwidth}
    \centering
    \includegraphics[width=\linewidth]{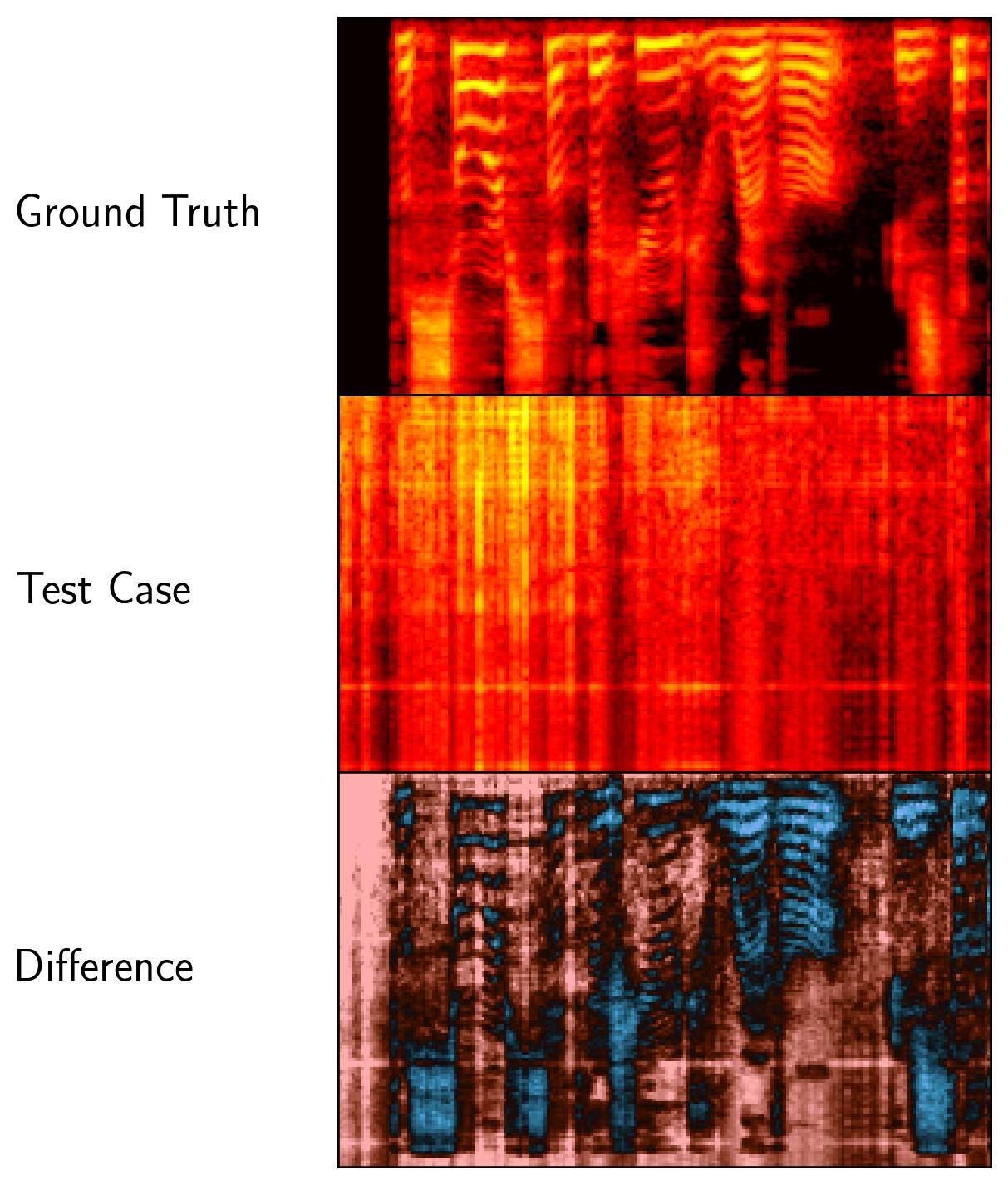}
    \caption{\wv}
\end{subfigure}
\hfill
\begin{subfigure}[t]{0.24\textwidth}
    \centering
    \includegraphics[width=\linewidth]{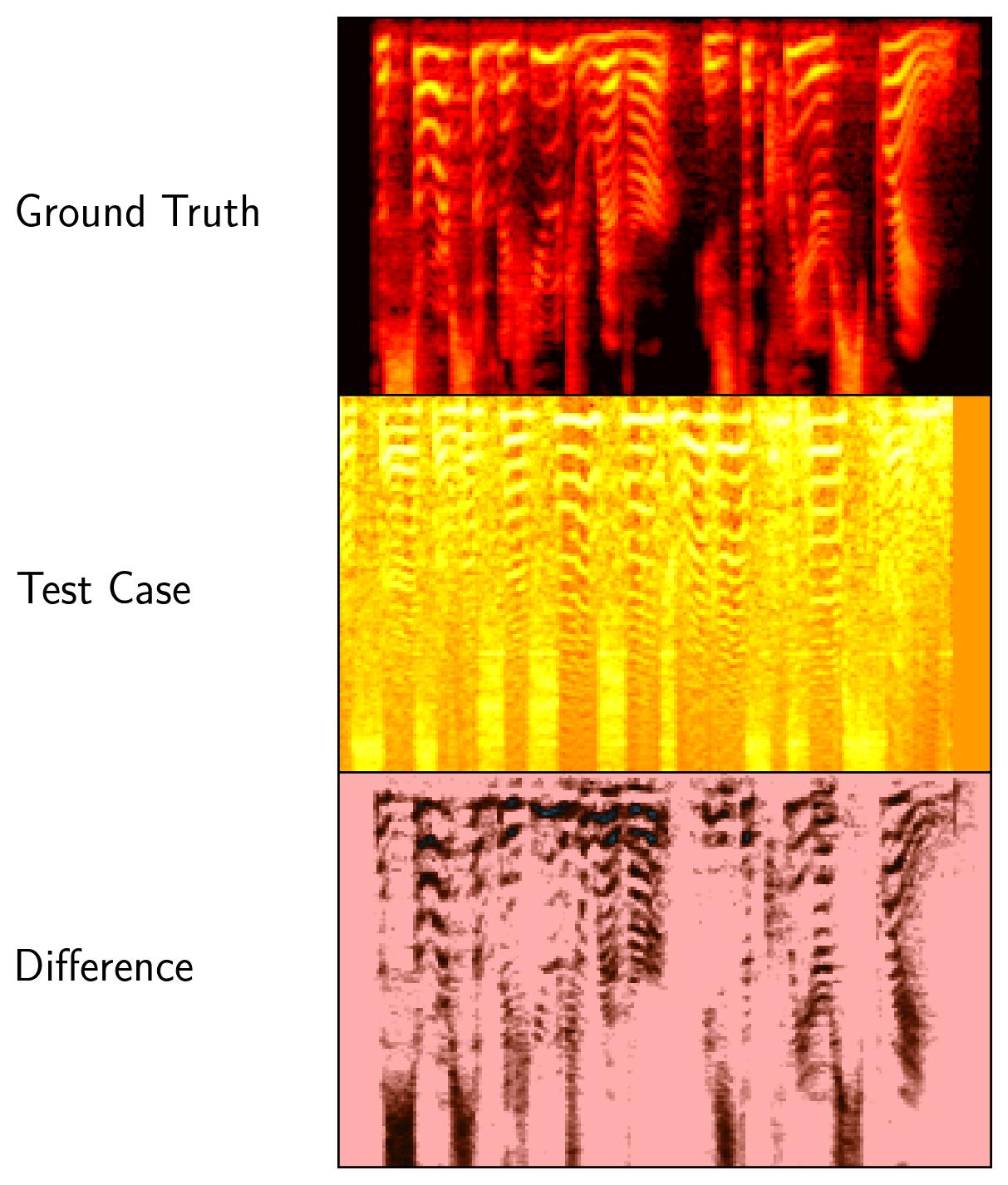}
    \caption{\smack}
\end{subfigure}
\hfill
\begin{subfigure}[t]{0.24\textwidth}
    \centering
    \includegraphics[width=\linewidth]{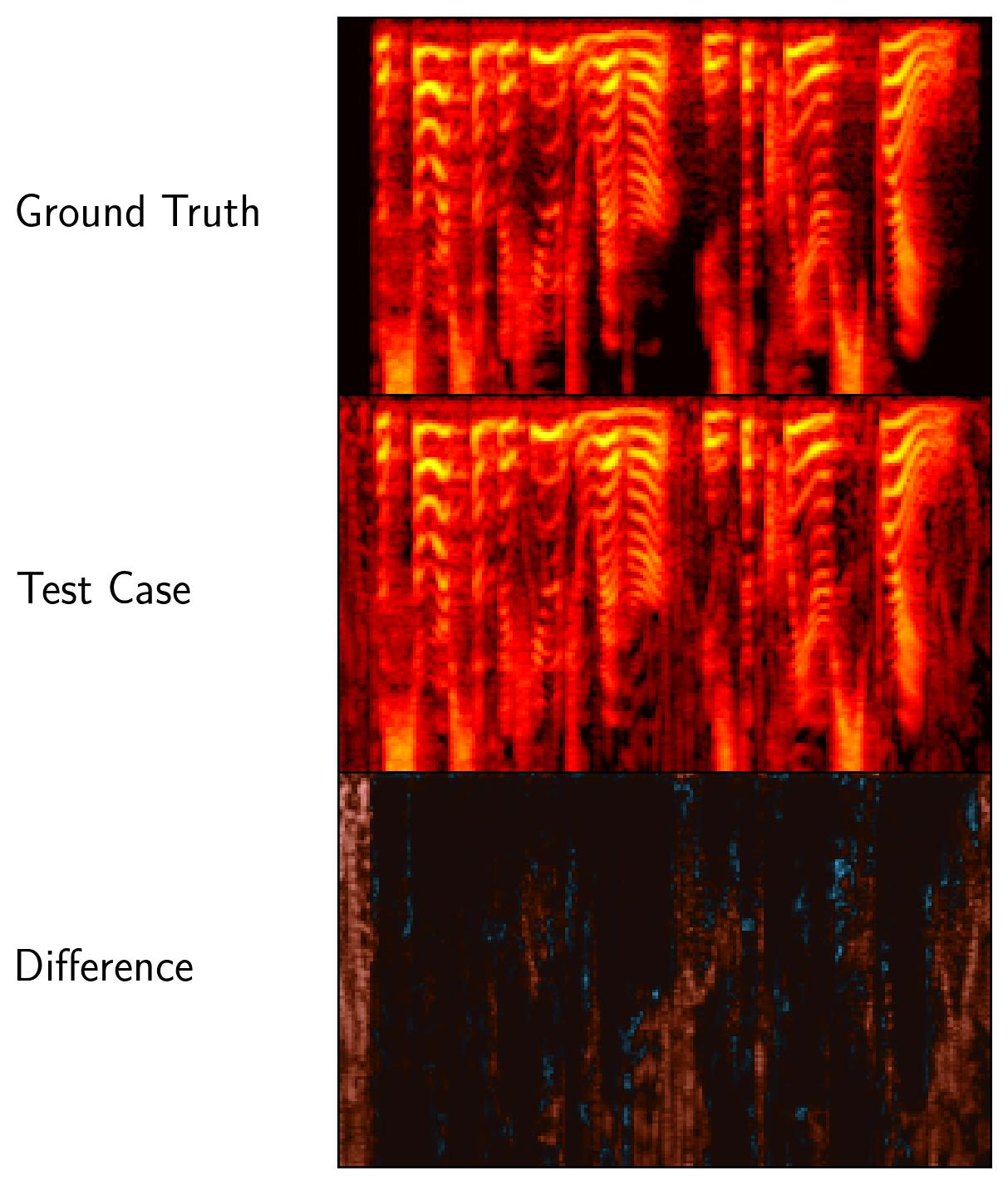}
    \caption{\pgd}
\end{subfigure}

\caption{Spectrogram comparison of the original audio versus the generated test case for each approach.}
\label{fig:three_subfigs}
\end{figure*}

\autoref{tab:performance} summarizes the performance of all methods across attack success, audio quality, and distortion-based metrics. Overall, \tool achieves a success rate of $0.980$, comparable to the baselines (all $\geq 0.990$), indicating that it maintains competitive effectiveness while optimizing other objectives.
In particular, looking at audio quality, \tool achieves the highest UTMOS score ($4.466$), slightly outperforming \pgd ($4.443$) and substantially exceeding \wv and \smack (both around $2.1$). This indicates that, even when failure exposure is comparable, \tool preserves naturalness significantly better than most baselines.

Across all distortion metrics, \tool consistently yields the lowest values. Specifically, it achieves a $d_F$-Spec of $0.017$, compared to at least $0.045$ for competing methods. Similarly, for $d_C$-Raw and $d_C$-Emb, \tool obtains $0.047$ and $0.439$, respectively, both substantially lower than all baselines. These results demonstrate that \tool introduces minimal perturbations in both signal space and embedding space.

The statistical analysis in ~\autoref{tab:stats} confirms that these improvements are significant. We compute $p$-values using the Wilcoxon signed-rank test~\cite{Wilcoxon1945}, and consider results to be statistically significant when $p < 10^{-2}$. For effect sizes we use Cohen's $d$, where values above $0.5$ indicate medium effect size and values above $0.8$ indicate large effect size.
Under this criterion, all comparisons against \wv and \smack are highly significant ($p < 10^{-15}$) with large effect sizes across all metrics, indicating substantial practical improvements. When compared to \pgd, the differences remain statistically significant across all metrics (e.g., $p = 4.44 \times 10^{-3}$ for UTMOS), although effect sizes are smaller (e.g., $d = 0.3$ for UTMOS), reflecting closer performance between the two methods.

\paragraph{Qualitative Analysis of Failures}
\autoref{fig:three_subfigs} provides qualitative examples. For each approach, the ground truth audio is shown on top, the generated test case in the middle, and the spectrogram difference at the bottom, for which blue regions is a negative change in dB in the final spectrogram, whereas red regions indicate a positive change in dB in the final testcase.

The visual differences align with the quantitative findings. \tool exhibits consistently low-intensity difference patterns, indicating minimal perturbations, while the generated audio remains visually close to the original despite inducing transcription changes. In contrast, \wv produces visibly stronger and more widespread modifications, and \smack introduces substantial structural changes, including altered signal length and broadly increased amplitudes. \pgd shows more localized but still noticeable modifications, often amplifying previously inactive frequency regions. These observations further support that \tool achieves lower distortion while maintaining effectiveness.

Beyond aggregate metrics, we analyze the nature of failures induced by \tool to better understand how perturbations affect ASR behavior. We observe that generated test cases do not produce uniform errors, but instead induce a range of qualitatively distinct failure modes.

First, we observe \emph{phonetic substitutions}, where specific words are replaced by acoustically similar alternatives (e.g., ``light'' $\rightarrow$ ``night''). These errors often preserve the overall sentence structure while altering the meaning, indicating that small changes in phonetic realization can shift decoding decisions.
One example is the sentence  
\textit{``This is a grand season for hikes on the road.''}  
which is transcribed by the ASR system as  
{\textit{``This is a grand Caesar for Heiks on the road.''} 
Although the transcription differs, the audio itself has not changed significantly from a human perception standpoint. This is illustrated in \autoref{fig:abs ex1}, which shows the difference in dB across frequency bands is quite isolated to one region in the middle explaining the change from \emph{hikes} to \emph{Heiks}, while \emph{season} to \emph{Cesar} required less change as they are pronounced quite similarly.

\begin{figure}
    \centering
    \includegraphics[width=\columnwidth]{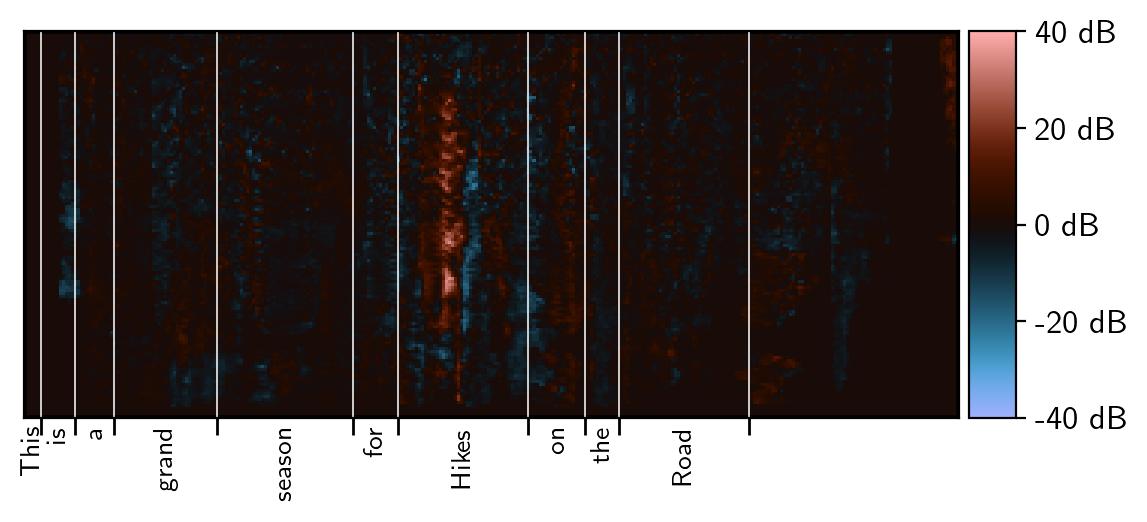}
    \caption{Example change in frequency band amplitudes.}
    \label{fig:abs ex1}
\end{figure}

Second, we identify \emph{semantic drift}, where multiple words are modified, resulting in a transcription that remains syntactically valid but semantically different from the original utterance. In these cases, errors accumulate across the sequence, suggesting that local perturbations can propagate through the model's decoding process. As an example the sentence \textit{``The boy was there when the sun rose``} changed to \textit{``The people who was there were the sunrobes``}.

Third, we observe \emph{partial degradation}, where only segments of the utterance are affected, leading to insertions, deletions, or localized substitutions, such as the case where \textit{``A pound of sugar costs more than eggs``} degrades to \textit{``A town of Shubber costs more than eggs``}.

A key distinction emerges when comparing these failures to waveform-based approaches. While both methods achieve high attack success, waveform perturbations often lead to near-complete transcription breakdown, as reflected by extremely high character error rates. In contrast, \tool maintains substantially lower error rates, indicating that the generated audio remains largely intelligible despite inducing incorrect outputs.

This suggests that \tool produces controlled deviations from the original transcription rather than global corruption. Instead of collapsing into noise, the system generates plausible alternative transcriptions that preserve structure while altering content. These differences stem from the structure of the search space: waveform-based methods frequently push inputs outside the natural speech manifold, whereas latent-space manipulation remains within it, enabling structured and linguistically plausible variations.

From a testing perspective, this is critical: \tool exposes diverse and realistic failure behaviors, while baseline approaches predominantly reveal sensitivity to signal degradation.

\begin{tcolorbox}[boxrule=0pt,sharp corners,boxsep=2pt,left=2pt,right=2pt,top=2.5pt,bottom=2pt]
\begin{center}
\begin{minipage}[t]{0.99\linewidth}
\textbf{RQ\textsubscript{1} (effectiveness):}~\tool maintains competitive attack success while achieving the highest audio quality and lowest distortion across all evaluated metrics. It therefore provides the most favorable trade-off between effectiveness and imperceptibility among the compared methods.
These results suggest that effectiveness is largely determined by the structure of the search space rather than the optimization algorithm. Latent-space manipulation achieves better trade-offs than waveform-based approaches, highlighting the importance of semantically aligned representations.
\end{minipage}
\end{center}
\end{tcolorbox}

\subsubsection{Validity (RQ\textsubscript{2})}
\label{sec:rq2}
\begin{table}[t]
\caption{Mean Opinion Scores on audio quality (RQ\textsubscript{2}).}
\label{tab:MOS}
\centering
\resizebox{0.9\columnwidth}{!}{%
\begin{tabular}{lcc}
\toprule
     & MOS Technical & MOS Semantic \\
\midrule
    \textit{Ground Truth} & $4.92 \pm 0.31$ & $4.01 \pm 1.24$ \\
    \tool & $\mathbf{4.81 \pm 0.54}$ & $3.86 \pm 1.26$ \\
    \wv & $1.24 \pm 0.53$ & $1.33 \pm 0.79$ \\
    \smack & $1.31 \pm 0.54$ & $1.57 \pm 0.78$ \\
    \pgd & $3.73 \pm 1.23$ & $\mathbf{3.98 \pm 1.30}$ \\
\bottomrule
\end{tabular}
}
\end{table}

\autoref{tab:MOS} reports MOS reported by human assessors for both technical and semantic audio quality. The \textit{ground truth} establishes an upper bound for technical quality ($4.92$) and a reference level for semantic consistency ($4.01$).

\tool achieves the highest technical MOS among all methods ($4.81$), closely approaching ground truth quality, indicating that perturbations have minimal impact on perceptual audio fidelity. In contrast, \wv and \smack show substantially degraded technical quality (both $\approx 1.3$), reflecting strong perceptual distortions. \pgd performs moderately ($3.73$), but remains noticeably below \tool.

For semantic MOS, \pgd attains the highest score ($3.98$), closely matching the ground truth, while \tool ranks second ($3.86$), indicating that semantic content is largely preserved. In contrast, \wv and \smack again perform poorly (both $\leq 1.57$), suggesting significant degradation of perceived content.

\begin{tcolorbox}[boxrule=0pt,sharp corners,boxsep=2pt,left=2pt,right=2pt,top=2.5pt,bottom=2pt]
\begin{center}
\begin{minipage}[t]{0.99\linewidth}
\textbf{RQ\textsubscript{2} (validity):}~\tool achieves the highest technical audio quality while maintaining competitive semantic fidelity, second only to \pgd. In contrast, \wv and \smack substantially degrade both technical and semantic quality. Overall, \tool provides the most balanced preservation of perceptual audio characteristics among the evaluated methods.
\end{minipage}
\end{center}
\end{tcolorbox}

\subsubsection{Efficiency (RQ\textsubscript{3})}
\label{sec:rq3}

\autoref{tab:efficiency} summarizes runtime and query budget across all methods. Since all approaches query the same system under test, the cost per query is identical across methods. Consequently, differences in runtime are attributable to algorithmic characteristics, such as optimization strategy and associated overhead.

As shown in \autoref{tab:efficiency}, \pgd achieves the lowest runtime ($15.86$ seconds) and the smallest query budget (200), reflecting its gradient-based optimization and direct access to model internals. Similarly, \smack operates with a fixed and low query budget (300) and a runtime comparable to \tool ($223.73$ seconds vs. $222.43$ seconds), indicating efficient execution under its specific optimization design.

In contrast, \tool and \wv exhibit substantially higher mean query budgets, with $5067.33$ and $7039.00$, respectively, as reported in \autoref{tab:efficiency}. Despite this, \tool achieves a lower average runtime ($222.43$ seconds) than \wv ($361.69$ seconds), indicating more efficient optimization dynamics given comparable query costs. The higher variance observed for both methods reflects the adaptive nature of the search process.

For \tool and \wv, early stopping criteria are applied as described in \autoref{sec:setup}, reducing unnecessary computation once suitable solutions are found. In contrast, early stopping is not applied to \smack and \pgd, as their optimization procedures differ and do not support comparable stopping conditions.

To assess convergence behavior and the impact of early stopping for \wv and \tool, we plot the cumulative success rate across test cases in \autoref{fig:cumsuc}. The results show that \tool exhibits a steeper increase, indicating that the early stopping criteria are met more frequently and earlier during optimization. In contrast, \wv shows a plateau after approximately 20 generations, suggesting difficulty in satisfying the defined thresholds.

Towards the end of the optimization process, both methods exhibit an increase in success rate. This is explained by the definition of success as any change in ASR transcription: even if qualitative thresholds are not met, termination of the optimization can still yield a successful outcome under this criterion.

\begin{table}[t]
\caption{Efficiency results (RQ\textsubscript{3}).}
\resizebox{0.9\columnwidth}{!}{%
\begin{tabular}{lllllll}\toprule
         & Runtime (sec) $\downarrow$   & Budget $\downarrow$   \\ \midrule
GATAS    & 222.43 $\pm$ 168.00 & 5067.33 $\pm$ 3752.31  \\
SMACK    & 223.73 $\pm$ 48.80                &   300.00 $\pm$ 0.00                 \\
Waveform & 361.69 $\pm$ 285.16 & 7039.00 $\pm$ 4151.12 \\
PGD      & \textbf{15.86 $\pm$ 1.27}    & \textbf{200.00. $\pm$ 0.00}    \\\bottomrule
\end{tabular}\label{tab:efficiency}
}
\end{table}

\begin{tcolorbox}[boxrule=0pt,sharp corners,boxsep=2pt,left=2pt,right=2pt,top=2.5pt,bottom=2pt]
\begin{center}
\begin{minipage}[t]{0.99\linewidth}
\textbf{RQ\textsubscript{3} (efficiency):}~\tool achieves competitive runtime while operating under substantially higher query budgets than fixed-budget baselines, indicating more efficient optimization dynamics compared to \wv. While gradient-based methods (\pgd) and constrained approaches (\smack) are faster due to lower query requirements, \tool benefits from early stopping and converges more reliably, reaching successful solutions earlier in the optimization process.
\end{minipage}
\end{center}
\end{tcolorbox}

\subsection{Threats to Validity}
\label{sec:threats}

\subsubsection{Internal Validity}

All methods are evaluated under identical optimization settings where applicable, including the same hyper-parameters and configurations, such as fitness functions, success thresholds, and query budgets, ensuring a fair comparison between the proposed method and the waveform baseline. Early stopping is applied consistently across all runs. 
However, exact parameter alignment is not always meaningful due to fundamental differences in method design and optimization requirements. In such cases, deviations are explicitly documented and justified to mitigate threats to internal validity.

The ASR system is treated as deterministic by enforcing greedy decoding; however, minor non-determinism may still arise from GPU execution (e.g., cuBLAS), potentially affecting reproducibility of individual runs. In the human study, variability in participant responses is inherent; this is addressed through randomized assignment, balanced condition exposure, and data cleaning to remove inconsistent or invalid responses.

\subsubsection{External Validity}

The evaluation is conducted on a single ASR system (Whisper tiny) and a single-speaker TTS model, which may limit generalizability to other architectures, larger ASR models, or multi-speaker and more diverse speech settings. The dataset consists of phonetically balanced but relatively short and semantically simple sentences, which may not fully represent real-world speech. Generalization to other languages, domains, and ASR systems is left for future work.

\subsubsection{Replicability}

To facilitate replication and enable public scrutiny of our results, we provide a replication package~\cite{replication-package} that includes the \tool implementation, experimental scripts, and datasets.

\begin{figure}[t]
    \centering
    \includegraphics[width=0.8\columnwidth]{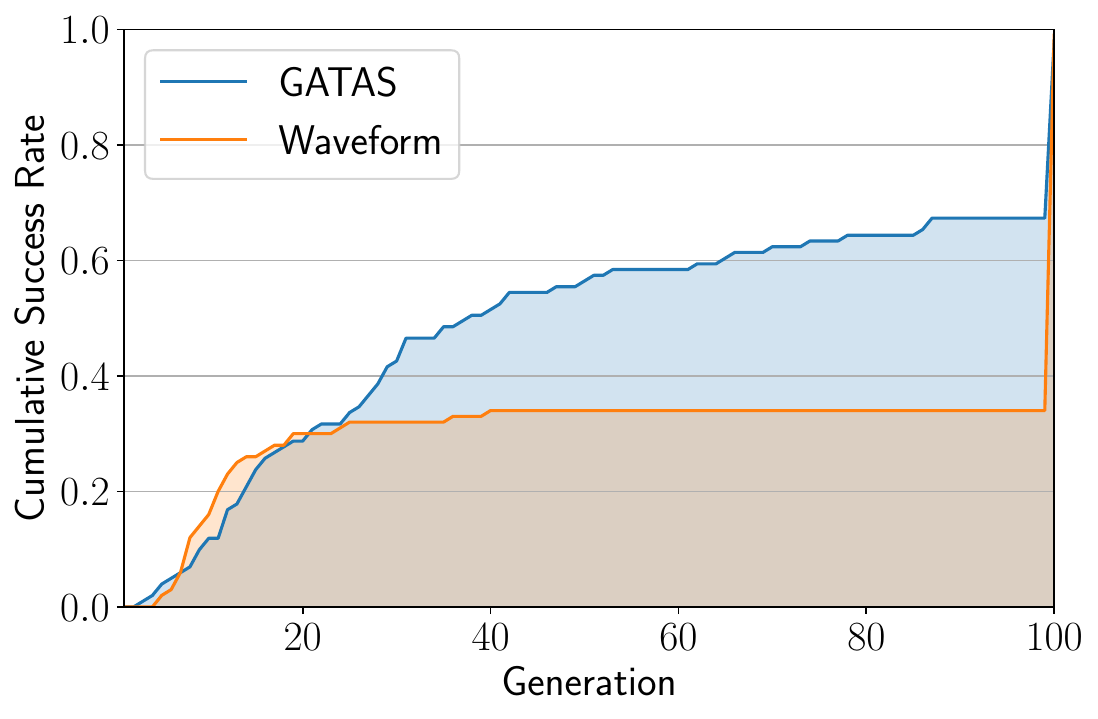}
    \caption{Cumulative success rate (RQ\textsubscript{3}).}
    \label{fig:cumsuc}
\end{figure}

\section{Discussion}
\label{chap:discussion}

\head{The performance gap between methods is structural, not algorithmic}
\tool and \wv use identical optimization setups, yet \tool achieves a 98\% success rate while significantly outperforming all baselines in quality and distortion metrics (\autoref{tab:performance}). This gap stems from the interpolation domain: in the TTS embedding space, each phoneme is controlled independently via a 512-dimensional vector, allowing semantic divergence and audio quality to improve simultaneously, as the TTS decoder constrains outputs to remain natural. In waveform space, semantic change requires signal degradation, making the two objectives antagonistic.

Human evaluation confirms this structural advantage. As shown in \autoref{tab:MOS}, \tool achieves the highest technical quality and the second-highest semantic quality, closely approaching ground truth audio. \smack, despite operating on prosodic feature representations rather than raw waveforms, was designed for targeted attacks against pre-transformer ASR systems; adapting it to an untargeted setting on Whisper produces results comparable to the \wv baseline. \pgd achieves comparable effectiveness and semantic fidelity to \tool, as expected given full gradient access, but requires white-box access unavailable in deployed systems and \tool provides substantially better audio quality and lower distortion even compared to this white-box method.

In terms of efficiency, \pgd achieves the lowest runtime due to gradient-based optimization, but \tool attains the second-lowest runtime despite a substantially larger query budget. This suggests that the well-conditioned embedding space allows \tool to convert search effort into successful solutions more effectively than other black-box approaches.

\head{Untargeted formulation is better suited to the speech domain}
Many black-box ASR attacks rely on predefined target transcriptions~\cite{yu2023smack, taori2019targeted, alzantot2018did, du2020sirenattack}, requiring the optimization to align generated audio with a specific output. In the speech domain, variability in phoneme duration and temporal structure makes convergence toward arbitrary target phrases difficult, often requiring larger perturbations that degrade naturalness. \tool sidesteps this by adopting an untargeted formulation that seeks any semantically divergent transcription rather than a specific one, relaxing alignment requirements while still enforcing quality constraints through the multi-objective optimization.

This distinction also explains \smack's weak performance in our evaluation. Originally designed for targeted optimization, its search strategy is tightly coupled to the presence of explicit targets; removing them degrades both effectiveness and audio quality.

\section{Related Work}
\label{sec:related_work}

Machine learning systems are increasingly evaluated through testing methods that probe model behavior under distribution shifts and structured input modifications. In the audio domain, such testing is constrained by human perception, requiring that modified inputs remain natural and intelligible. This section reviews prior work on generative testing approaches and ASR-specific methods under different access models, highlighting the absence of naturalness-preserving black-box testing strategies.

\subsection{Generative Testing Approaches}

Generative models provide a mechanism for testing model behavior by producing inputs directly from the data distribution. Rather than relying on input-level perturbations, these methods operate in latent spaces, where learned representations constrain outputs to remain realistic while enabling structured exploration.

Early work primarily relied on GANs. Zhao et al.~\cite{zhao2017generating} showed that manipulating GAN latent representations yields realistic inputs that can expose model failures, while Song et al.~\cite{song2018constructing} introduced constraints to balance realism and misclassification. Subsequent approaches demonstrated that structured latent spaces enable controlled exploration of the data manifold while preserving perceptual validity~\cite{weissl2025targeted, merabishvili2026latent, 2026-Chen-DETECT}.

Latent-space manipulation has since been explored across different generative model classes. Noise-based methods perturb latent representations in VAEs~\cite{kang2020sinvad}, GANs~\cite{buzhinsky2023metrics}, and diffusion models~\cite{maryam2025benchmarking}. While these approaches enable broad exploration, they typically lack precise control over the resulting samples. Diffusion models have recently gained prominence due to their higher fidelity and stability~\cite{maryam2025benchmarking, missaoui2023semantic, bao2024generative, weissl2026hypernet}, and have been used for tasks such as semantically guided input synthesis via inpainting~\cite{missaoui2023semantic}. However, even in diffusion-based approaches, manipulation is often indirect and constrained by the underlying generative process.

In audio, generative testing remains comparatively underexplored, especially in the audio domain. Xie et al.~\cite{xie2021enabling} propose a model that generates perturbations in a single forward pass, improving efficiency but still operating at the signal level and targeting classification tasks. Cheng et al.~\cite{cheng2024alif} manipulate linguistic embeddings within a TTS model, but explicitly maximize distortion, resulting in outputs that are not intelligible to humans.

Overall, prior work shows that latent-space manipulation can preserve naturalness by construction while enabling systematic test generation. However, existing approaches either rely on weakly controlled perturbations or do not enforce perceptual quality. This gap is particularly evident for ASR testing in black-box settings.

\subsection{White-box Testing of ASR Systems}

White-box testing methods assume access to model internals such as gradients and alignments. Carlini and Wagner~\cite{carlini2018audio} demonstrated that ASR systems can be systematically probed by optimizing inputs to produce specific transcriptions, exploiting sequence alignment mechanisms. Schönherr et al.~\cite{schonherr2018adversarial} incorporated psychoacoustic masking to ensure that modifications remain inaudible by constraining them to masked frequency regions.

These approaches enable precise and effective testing of ASR behavior, but rely on full model access and are therefore not applicable to deployed systems accessed via APIs.

\subsection{Black-box Testing of ASR Systems}

Black-box testing methods operate without access to gradients and instead rely on model queries or transferability. A large body of work explores search-based strategies in waveform space. Alzantot et al.~\cite{alzantot2018did} used genetic algorithms to evolve audio inputs that alter ASR outputs, demonstrating feasibility in a black-box setting. Taori et al.~\cite{taori2019targeted} extended this to full sentences using hybrid optimization, while Du et al.~\cite{du2020sirenattack} applied particle swarm optimization. These approaches consistently operate directly on the waveform, leading to perceptible artifacts.

Subsequent work improves efficiency and search strategies. Occam~\cite{zheng2021black} reduces dimensionality via segmentation, though it assumes limited cross-segment dependencies. Kenansville~\cite{abdullah2021hear} removes frequency components without iterative optimization, trading effectiveness for audio degradation. NP-Attack~\cite{biolkova2022neural} models the decision boundary explicitly to guide search under hard-label constraints, but remains waveform-based. Other works use a differential testing approach to uncover failures in ASR systems, however focusing on manipulations only in the text sequence space~\cite{asyrofi2020crossasr, asyrofi2021crossasr++}.

Transfer-based approaches optimize inputs against surrogate models. Devil's Whisper~\cite{chen2020devil} constructs substitute models using TTS-generated data and improves transfer through ensembling, while Fang et al.~\cite{fang2024zero} optimize across multiple models without querying the target system. These methods depend on transferability assumptions and do not explicitly control perceptual quality.

Specialized approaches exploit domain-specific properties. Ultrasonic methods such as DolphinAttack~\cite{zhang2017dolphinattack} and UltraAdv~\cite{zhang2025ultraadv} operate outside the human hearing range but require specific hardware conditions. Other methods target model-specific behaviors, such as Whisper muting attacks~\cite{raina2024muting}. These approaches are effective in constrained scenarios but lack general applicability.

Across black-box methods, a consistent limitation emerges: operating in waveform space makes it difficult to simultaneously explore model behavior and preserve perceptual naturalness, which is one of the design goals behind \tool.

\section{Conclusions and Future Work}\label{sec:conclusion}

This work introduced \tool, a black-box method for generating test cases for ASR systems by operating in the phoneme-level embedding space of a TTS model. The results show that this representation enables simultaneous control over transcription changes and audio quality, producing test cases that remain natural while reliably inducing misrecognition.
Compared to waveform-based methods, phoneme-level latent manipulation yields lower distortion and higher audio quality at comparable success rates. In addition, \tool achieves performance close to a white-box baseline without requiring access to model internals, indicating that representation choice and objective design are more critical than gradient access for this task.
The use of an untargeted formulation simplifies test case generation by removing the need to define target transcriptions, while still enabling the discovery of diverse failure cases under perceptual constraints.

Several directions can further extend this approach. One option is to incorporate surrogate models that guide the search toward semantically meaningful regions in phoneme-level embedding space, enabling more controlled exploration. Another direction is to study how latent trajectories relate to semantic changes in transcription, providing a more structured way to steer test case generation. 
While our approach focuses on inducing transcription errors under perceptual constraints, real-world deployments often involve downstream components such as dialogue managers or decision systems, where the impact of misrecognition may be amplified or mitigated. Investigating how latent-space perturbations propagate through such pipelines would enable a more comprehensive assessment of system robustness. Finally, extending the method to other speech models and tasks would allow assessing how well the approach generalizes beyond the evaluated setting.

\section{Data Availability Statement}

All code and data required to reproduce the results are provided in our anonymized replication package~\cite{replication-package}, including implementation, experiment scripts, and raw human evaluation data.

\balance
\bibliographystyle{ACM-Reference-Format}
\bibliography{bibliography}

\end{document}